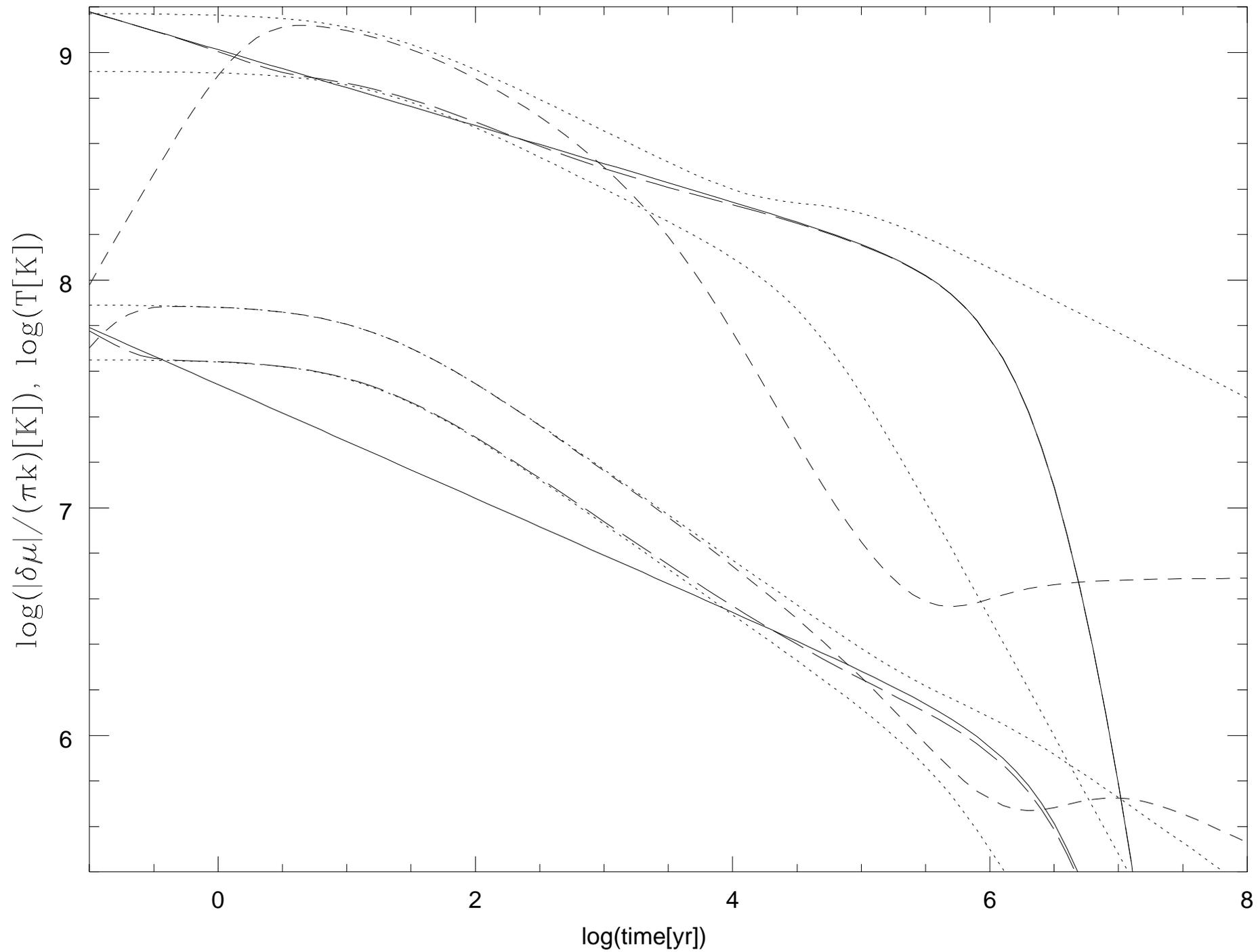

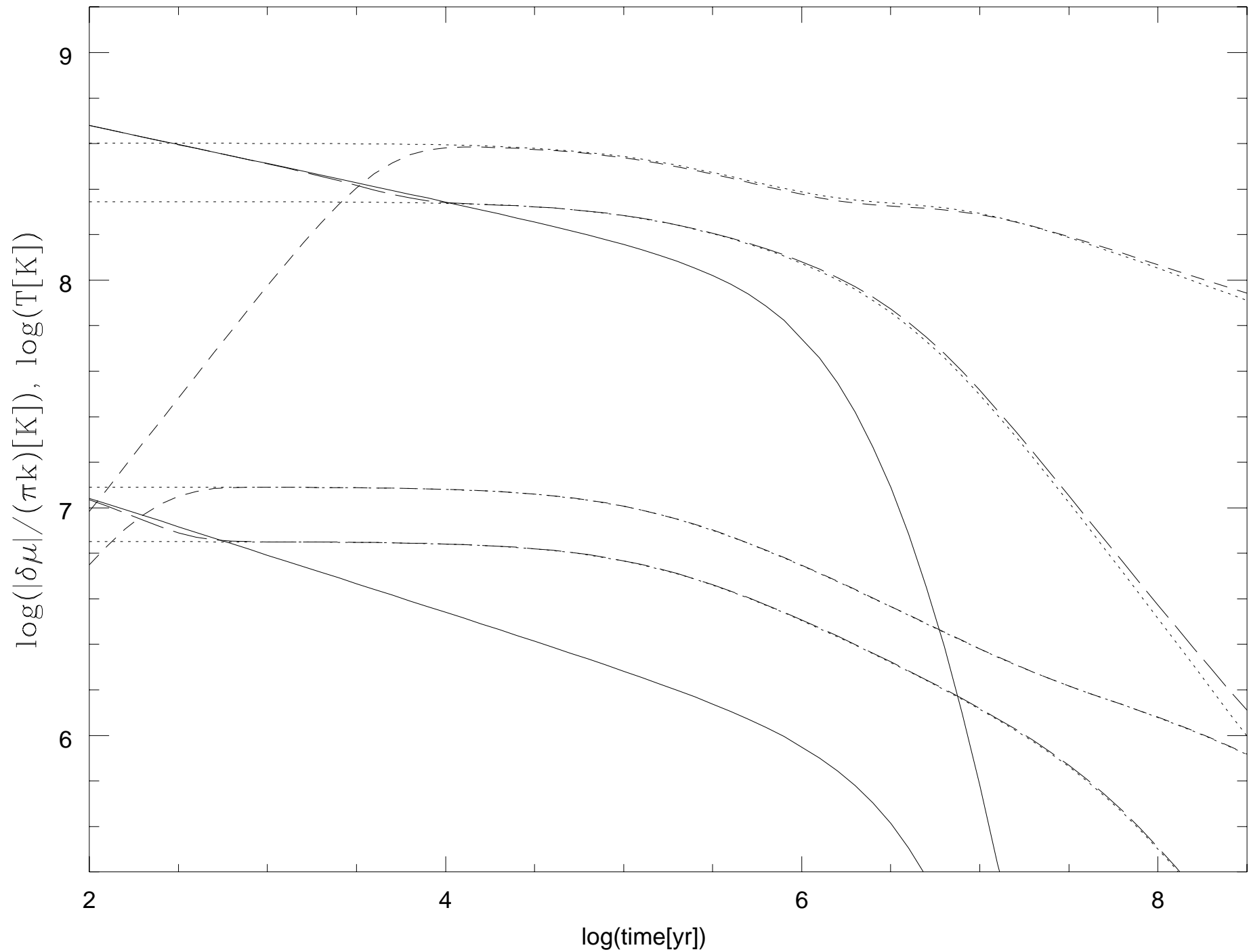

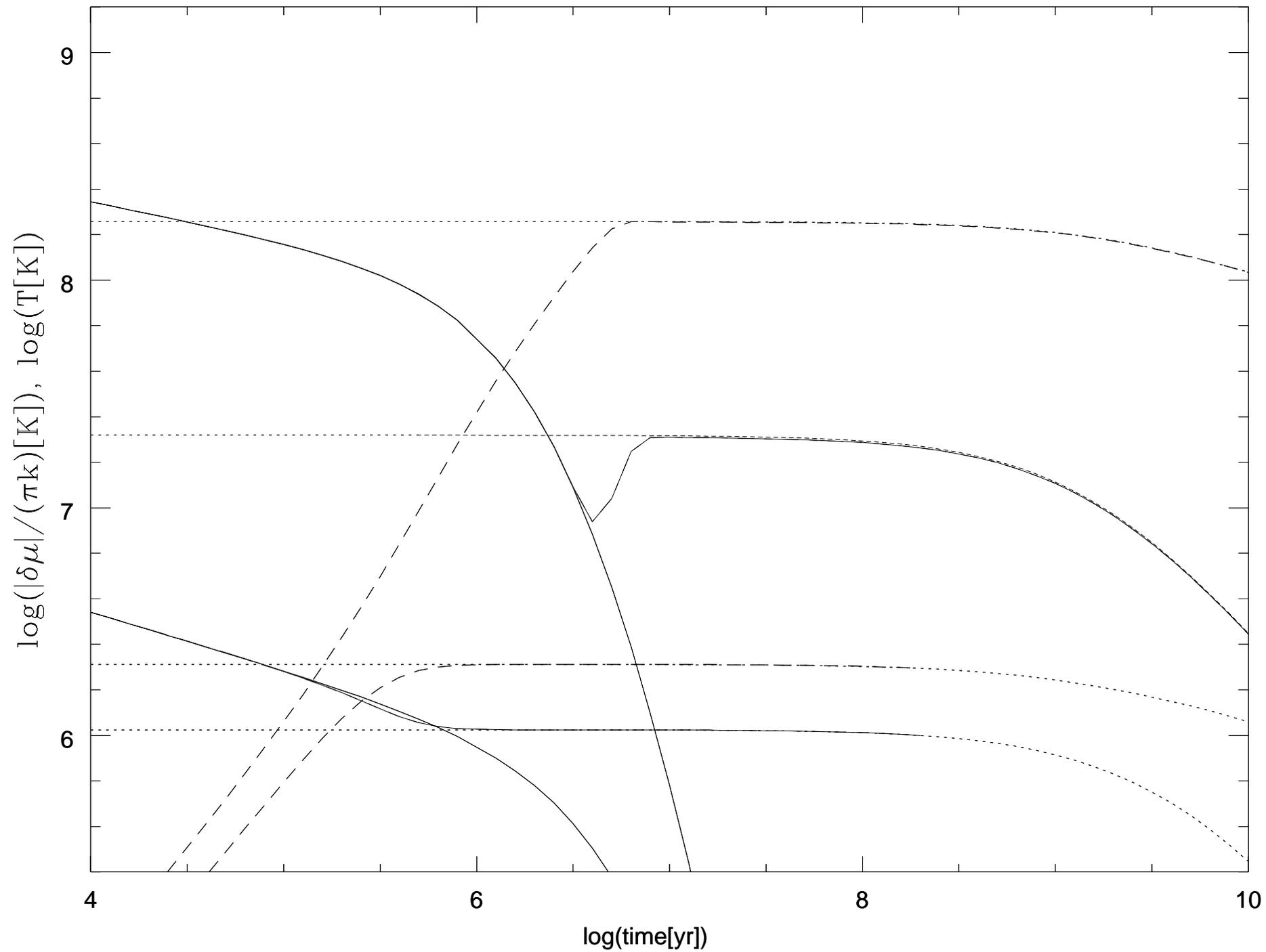

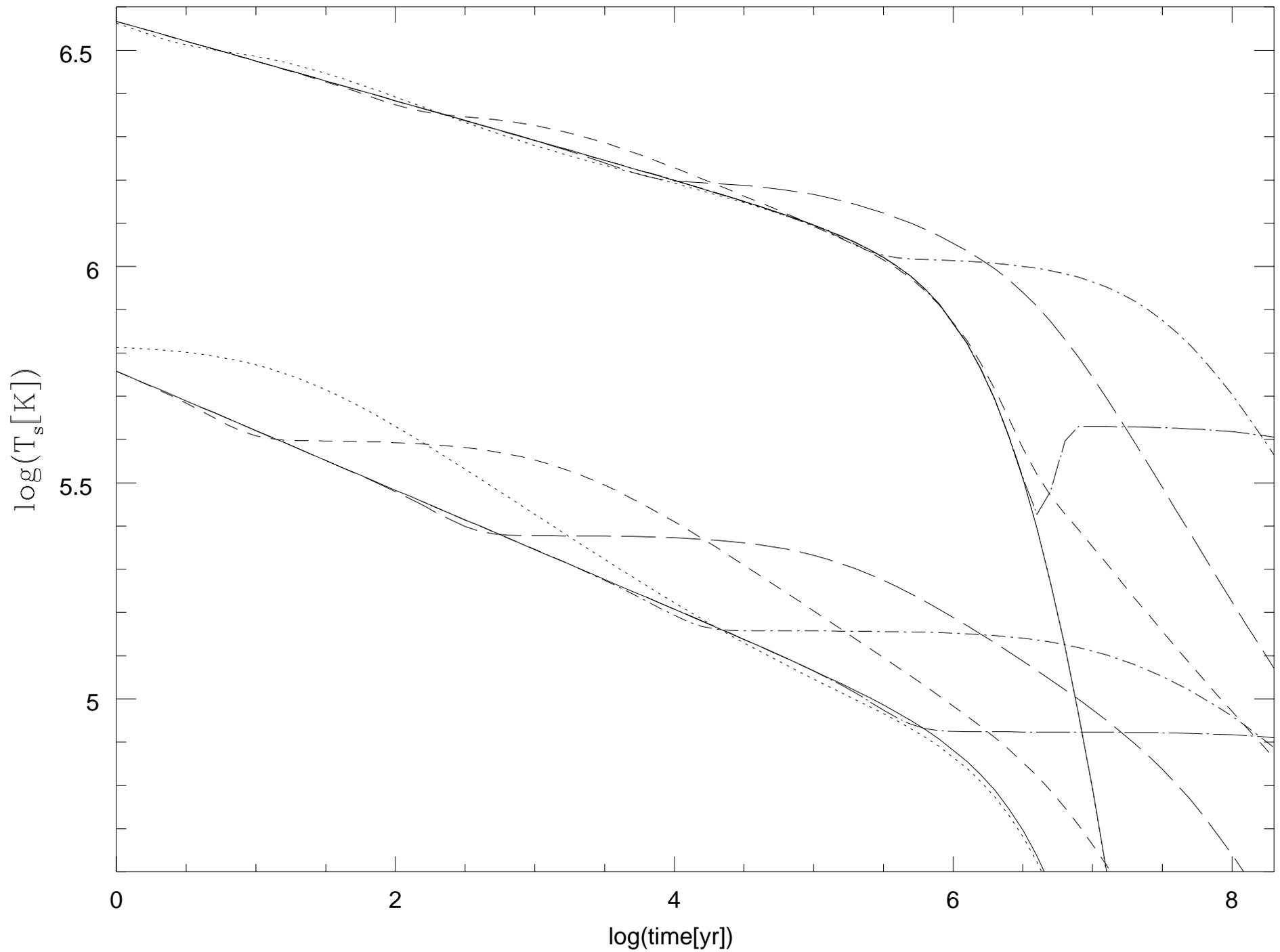

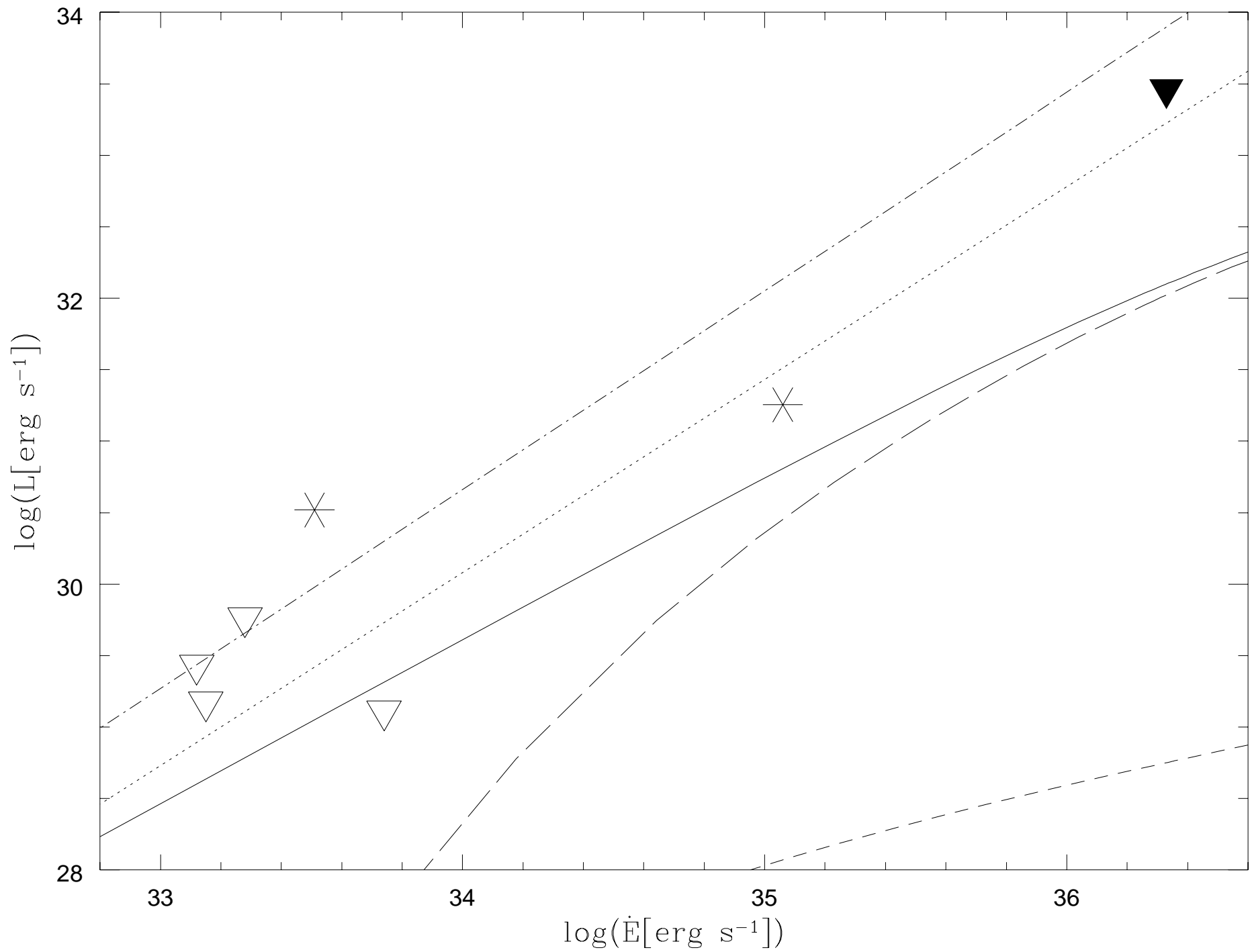

# DEVIATIONS FROM CHEMICAL EQUILIBRIUM
# DUE TO SPIN-DOWN
# AS AN INTERNAL HEAT SOURCE IN NEUTRON STARS

Andreas Reisenegger

Institute for Advanced Study, Princeton, NJ 08540

Electronic mail: andreas@guinness.ias.edu

## ABSTRACT

The core of a neutron star contains several species of particles, whose relative equilibrium concentrations are determined by the local density. As the star spins down, its centrifugal force decreases continuously, and the star contracts. The density of any given fluid element increases, changing its chemical equilibrium state. The relaxation towards the new equilibrium takes a finite time, so the matter is not quite in chemical equilibrium, and energy is stored that can be released by reactions. For a neutron star core composed of neutrons (n), protons (p), and electrons (e), the departure from chemical equilibrium is quantified by the chemical potential difference $\delta\mu \equiv \mu_{\rm p} + \mu_{\rm e} - \mu_{\rm n}$. A finite $\delta\mu$ increases the reaction rates and the neutrino emissivity. If large enough ($|\delta\mu| \gtrsim 5kT$), it reduces the net cooling rate because some of the stored chemical energy is converted into thermal energy, and can even lead to net heating. A simple model (for non-superfluid matter) shows the effect of this heating mechanism on the thermal evolution of neutron stars. It is particularly noticeable for old, rapidly spinning stars with weak magnetic fields. If the timescale for variations of the rotation rate is much longer than the cooling time, a quasi-equilibrium state is reached in which heating and cooling balance each other and the temperature is completely determined by the current value of $\dot{P}/P^3$ (or the spin-down power). If only modified Urca reactions are allowed, the predicted quasi-equilibrium x-ray luminosity of some millisecond pulsars approaches the upper limits obtained by Danner et al. (1994) from ROSAT data. The predicted x-ray luminosity is much lower if direct Urca or other fast reactions are allowed. In both cases, the luminosity is probably increased if the stellar interior is mostly superfluid.

*Subject headings:* dense matter — equation of state — stars: neutron — stars: pulsars: general — stars: rotation — x-rays: stars

# 1. INTRODUCTION

The theoretical study of the thermal evolution of neutron stars and the comparison of its predictions to observations of the surface emission of neutron stars of known ages can (at least in principle) serve as an interesting and powerful diagnostic of the properties of matter at supernuclear densities. The main reason for this is that the dominant cooling mechanism of a hot neutron star is the emission of neutrinos and antineutrinos produced in the so-called Urca processes (Gamov & Schoenberg 1941):

$$\begin{aligned} \mathrm{n} &\longrightarrow \mathrm{p} + \mathrm{e} + \bar{\nu}_\mathrm{e}, \\ \mathrm{p} + \mathrm{e} &\longrightarrow \mathrm{n} + \nu_\mathrm{e}. \end{aligned} \quad (1)$$

The symbols n, p, e, $\nu_\mathrm{e}$, and $\bar{\nu}_\mathrm{e}$ denote neutrons, protons, electrons, electron neutrinos, and electron antineutrinos. The neutrino emissivity depends sensitively on the microphysics of the neutron star core. The reactions can proceed as written ("direct Urca reactions") only if the ratio $x = n_\mathrm{p}/n$ of the proton number density to the total baryon number density exceeds a certain threshold value which allows simultaneous energy and momentum conservation if all reacting particles are on the surface of their respective Fermi seas (Boguta 1981; Lattimer et al. 1991). Otherwise, an additional particle participates in the reaction and absorbs the additional momentum, without changing its "flavor" ("modified Urca reactions;" Chiu & Salpeter 1964). If the additional particle is another neutron or proton, the rate is strongly suppressed, because of large phase space reduction due to degeneracy. The suppression is smaller if the momentum can be absorbed by bosons such as pions (Bahcall & Wolf 1965) or kaons (Brown et al. 1988). These and other variants of Urca reactions are reviewed by Pethick (1992).

In addition to Urca reactions, other neutrino emitting processes (described in Nomoto & Tsuruta 1987) contribute to the cooling, as does the emission of thermal x-ray and ultraviolet photons from the stellar surface, which dominates for cooler stars. All cooling processes are strongly affected by Cooper pairing (superfluidity or superconductivity) of the baryons, which reduces the reaction rates and (at low enough temperatures) the heat capacity of the paired particles (e.g., Wolf 1966; Maxwell 1979; Levenfish & Yakovlev 1994a, b).

Some heating mechanisms may also be present. In isolated neutron stars, the most important source of energy that is likely to be tapped is the kinetic energy of rotation, which can be converted into internal heat in several ways. If the star does not rotate as a solid body, the energy in the differential rotation can be dissipated. This is predicted by the standard models of pulsar "glitches," which involve a crustal neutron superfluid with pinned vortex lines (Alpar et al. 1984; Shibazaki & Lamb 1989; Umeda et al. 1993; Van Riper, Link, & Epstein 1994). Differential rotation and the resulting dissipation may also occur in the core of the star, especially if the latter is stably stratified (Pethick 1991; Reisenegger & Goldreich 1992) and therefore cannot be spun up by the standard Ekman pumping process (Walin 1969; Sakurai 1969). The differential rotation may still be impeded by a strong magnetic field threading the core, but the existence of a magnetic field in the core is not well established, and the spin-down process in a strongly stratified, conducting fluid in the presence of a magnetic field has not yet been studied. In addition, as the star slows down, its equilibrium shape changes, and the response of the solid crust to this change, through cracking or plastic flow, also generates heat (see Cheng et al. 1992 and references therein).

The purpose of this paper is to point out another channel by which a neutron star's rotational energy can be converted into heat, and to study its effects. The physical idea is as follows. The core of the star contains several kinds of particles, whose relative concentrations adjust by Urca reactions. The equilibrium concentration of particles of each kind depends on density. As the star spins down, the centrifugal force decreases continuously, increasing the star's internal density. This changes the chemical equilibrium state throughout the core. Since the relaxation of the relative particle concentrations requires a finite amount of time, the star is never exactly in chemical equilibrium, so energy is stored. In addition, the finite departure from chemical equilibrium modifies the reaction rates (Haensel 1992). If the departure from equilibrium is large enough, the net effect of the reactions is to increase the thermal energy at the expense of the stored chemical energy.

The dissipation associated with reactions produced by density changes is often called "bulk viscosity" or "second viscosity" (Landau & Lifshitz 1959), and has been studied as a damping mechanism for oscillations and gravitational-wave instabilities of neutron stars (Finzi & Wolf 1968; Sawyer 1989; Cutler, Lindblom, & Splinter 1990; Haensel & Schaeffer 1992; Reisenegger & Goldreich 1992; Lai 1994)[1], and as a heating process in the collapse of a neutron star to a black hole (Gourgoulhon & Haensel 1993), but not (to my knowledge) in the context in which it is discussed in the present paper. I will avoid the terms "bulk viscosity" and "second viscosity," since they seem to imply a dissipation process that depends locally in time and in space on the divergence of the velocity field, as suggested, e.g., by equation (15.6) of Landau & Lifshitz (1959). Dissipation by shear viscosity does, in fact, depend locally on the shear of the velocity field (if the mean free path of the particles is much smaller than the macroscopic scales of interest), but "bulk viscosity" can be substantially more subtle and complicated, as should become clear below, particularly in §3.1.

The contraction of a neutron star due to spin-down also causes adiabatic heating. However, the adiabatic temperature change produced by a given density change $\Delta\rho$ is $\Delta T \sim (\Delta\rho/\rho)T$, much smaller than the change from the process proposed above, for which $\Delta T \sim (\Delta\rho/\rho)\mu_\mathrm{n}$, where $\mu_\mathrm{n}$ is the chemical potential (or Fermi energy) of the neutrons (not including rest mass). Even smaller are the effects of the contraction due to cooling (Baym 1981).

This paper is organized as follows. In §2, I briefly discuss the equilibrium properties of a rotating star composed of neutrons, protons, and electrons, and their dependence on the rotation rate, $\Omega$. In §3, I attempt a description of the chemical and thermal evolution of the star as $\Omega$ changes with time. In §3.1., I outline the (possibly very complicated) interplay between reactions and diffusive processes in stars with nontrivial internal structure, such as superfluid and non-superfluid regions, or regions with substantially different reaction rates. In the subsequent sections, the spatial structure of the star is ignored, except for the use of a standard formula relating the internal and surface temperatures. Section 3.2 presents a very simplified model for the chemical and thermal evolution of a non-superfluid star, which reduces to two coupled first-order differential equations for the typical values of the departure from chemical equilibrium $\delta\mu$ and the internal temperature $T$. The needed total neutrino and antineutrino emissivities and differential reaction rates are given in the Appendix. Section 3.3 contains the results of the numerical integration of the differential equations for stars spinning down due to magnetic dipole braking with different surface magnetic field strengths, and a qualitative analysis of these results. Changes to be expected with the introduction of pairing (superfluidity) are discussed in §3.4. In §4, I comment on the shortcomings of the present model, suggest possible improvements, compare the proposed mechanism to previously studied heating processes, and make a preliminary comparison of my results with recent observations.

---

[1] The reaction rates used by Finzi & Wolf (1968) are not correct (Haensel 1992). Lai (1994) incorrectly distinguishes between "bulk viscosity" and "damping by neutrino emission," which in fact correspond to the same physical process.

## 2. EQUILIBRIUM STATE OF A ROTATING STAR

I consider a neutron star that is spinning down (or up) at a rate slow enough that it can always be considered to be in hydrostatic equilibrium and in uniform (rather than differential) rotation. Because of the high degeneracy of the neutron star matter, its cooling time and its relaxation time towards chemical equilibrium are much longer than its dynamical timescale, and also may be longer than the characteristic spin-down time. For simplicity, I use Newtonian dynamics throughout this paper.

In hydrostatic equilibrium,

$$0 = -\frac{1}{\rho}\boldsymbol{\nabla}p - \boldsymbol{\nabla}\phi, \qquad (2)$$

where $\rho$ is the mass density, $p$ is the pressure, $\phi = \psi - (1/2)\Omega^2 r^2 \sin^2\theta$ is the effective gravitational potential (including the centrifugal potential), $\psi$ is the actual gravitational potential, $\Omega$ is the rotation rate, $r$ is the radial coordinate, and $\theta$ is the colatitude. Taking the curl of eq. (2), one sees that the surfaces of constant $\rho$, constant $p$, and constant $\phi$ coincide. It is convenient to label these surfaces by the amount of mass $M$ enclosed by each of them, as $\mathcal{S}(M)$, and in this way be able to identify surfaces at different rotation rates, although the values of $\rho$, $p$, and $\phi$ on $\mathcal{S}(M)$ change as the star spins down.

Since the matter in a neutron star core is composed of neutrons, protons, electrons, and other particles (muons, perhaps kaons, etc.), its equation of state must in general be written as $p = p(\rho, x_1, x_2, ...)$ where $x_i$ are the relative concentrations of different particle species. In this paper, I consider only neutrons, protons, and electrons. In this case, the proton number density $n_\mathrm{p}$ and the electron number density $n_\mathrm{e}$ are required to be equal in order to preserve charge neutrality. Thus, a single "composition parameter" $x = n_\mathrm{p}/n$ is enough to characterize the chemical state of the matter ($n = n_\mathrm{n} + n_\mathrm{p}$ is the total baryon number density, and $n_\mathrm{n}$ is the neutron number density). Since $\rho$ and $p$ are constant on each surface $\mathcal{S}(M)$, this must also be true for $x$. (If additional particles are present, more "composition parameters" become necessary, and these need not individually be constant on these surfaces in order for $p$ and $\rho$ to be constant simultaneously.)

The chemical equilibrium state of the matter at a given baryon density, $n$, and given entropy per baryon, $s$, is found by minimizing the total energy per baryon, $E(n, x, s)$ (including the contribution from the electrons), with respect to $x$, i.e., by demanding that

$$\frac{\partial E}{\partial x} = \delta\mu \equiv \mu_\mathrm{p} + \mu_\mathrm{e} - \mu_\mathrm{n} = 0, \tag{3}$$

where $\mu_\mathrm{n}$, $\mu_\mathrm{p}$, and $\mu_\mathrm{e}$ are the chemical potentials of the neutrons, protons, and electrons, respectively. The equilibrium value of $x$ ($x_\mathrm{eq}$) is a function of $n$.

A change in the stellar rotation rate changes the centrifugal force, and therefore the baryon density $n$ on a given surface $\mathcal{S}(M)$,

$$\frac{\Delta n}{n} \approx -\frac{\alpha\Omega\Delta\Omega}{G\rho_\mathrm{c}}. \tag{4}$$

for $\Omega^2 \ll G\rho_\mathrm{c}$. Here, the symbol $\Delta$ denotes the Lagrangian variation (i.e., the variation on a given surface $\mathcal{S}(M)$) of the quantity following it, $G$ is the gravitational constant, $\rho_\mathrm{c}$ is the central density of the star, and $\alpha$ is a positive, dimensionless number of order unity that depends only slightly on radius within the core of a neutron star. In the center of a star with the analytically solvable polytropic equation of state $p \propto \rho^2 \propto n^2$ (which falls into the range of theoretically allowed equations of state of neutron star matter), $\alpha = (\pi^2 - 3)/3\pi \approx 0.73$. Thus, the equilibrium value of $x$ varies as

$$\Delta x_\mathrm{eq} = \frac{dx_\mathrm{eq}}{dn}\Delta n = \alpha\frac{nE_{nx}}{E_{xx}}\frac{\Omega\Delta\Omega}{G\rho_\mathrm{c}}. \tag{5}$$

Here and in what follows, italicized subscripts on $E$ are used to denote partial derivatives of the energy per baryon (e.g., $E_{nx} \equiv \partial^2 E/\partial n\partial x$) to be evaluated in the cold, chemical equilibrium state ($T = \delta\mu = 0$).

## 3. CHEMICAL AND THERMAL EVOLUTION

### *3.1. Reactions vs. diffusion*

Two effects make the actual value of $x$ on $\mathcal{S}(M)$ change as $\Omega$ varies. First, since $x_\mathrm{eq}$ changes, the matter is out of chemical equilibrium, and this induces a difference $\Gamma$ between the rates (per unit volume) of neutron beta decays and inverse beta decays that goes in the direction of restoring the equilibrium. Second, the total chemical potentials (including internal and external contributions, cf. Kittel & Kroemer 1980) of the individual particle species typically become different on different surfaces $\mathcal{S}(M)$, thus deviating from diffusive equilibrium. If the timescale for diffusion of particles is shorter than the timescale for relaxation to chemical equilibrium, significant diffusion occurs and the relaxation to the chemical equilibrium state is a global process involving the whole stellar core. In the opposite case, relaxation proceeds locally, without significant interactions among different parts of the star.[2] Both regimes may occur in a real neutron star, depending on its history and on the uncertain properties of neutron star matter. The first is more likely if a superfluid is present (which inhibits reactions by reducing the available phase space, and probably also facilitates relative motions of the superfluid with respect to other particles), if the temperature is low, or if only modified Urca processes are allowed. The opposite conditions tend to favor the second regime.

For a given function $\Gamma(n, \delta\mu, T)$, the first regime is more efficient in restoring equilibrium, since most reactions can occur in those regions where $\Gamma$ is larger. For example, assume direct Urca reactions are allowed in the inner core region of the star but not in the outer core. If the diffusion time is short, $\delta\mu$ is always similar in both regions, with particles diffusing from one to the other. Reactions occur almost exclusively in the inner region, and most of the heat is generated there. If the diffusion time is long, the inner core always stays closer to chemical equilibrium than the outer core. Reactions occur at comparable rates in both regions, and most of the heat generation and energy emission (in the form of neutrinos) occurs in the outer region.

Gnedin and Yakovlev (1994) have proposed that a temperature gradient in the core of a neutron star may generate convective motions that would further complicate the thermal evolution. However, these authors do not take the strong stabilizing effect of the composition gradient into account, which wins by a large factor over the destabilizing effect of the temperature gradient (Pethick 1991; Reisenegger & Goldreich 1992), except possibly during the first few seconds after the supernova explosion (Thompson & Duncan 1993). The existence of a composition gradient also argues strongly against convection driven by a divergence of the magnetic stress tensor, which has been proposed by Urpin and Ray (1994). Radial motions of matter must be accompanied either by ambipolar diffusion (i.e., relative motions of different particle species) with relative velocities comparable to the bulk displacements, or by fast reactions that keep the moving fluid always close to its current chemical equilibrium state (Pethick 1991; Goldreich & Reisenegger 1992).

The model studied in the following sections ignores all these complications. It is unclear at this point how their inclusion would affect the predicted surface temperature of the star. This uncertainty should be kept in mind when analyzing the results of the calculations reported below. If possible, it should be addressed by future, more refined calculations.

### 3.2. A simple model

In this section, I present a simple model for the thermal evolution of a neutron star in the presence of internal chemical heating due to spin-down. I take the matter to consist of non-superfluid neutrons, protons, and electrons, and assume that the chemical and thermal state of the star can be described by just two variables, which are assumed to be independent of position within the star: the departure from chemical equilibrium $\delta\mu$, defined in eq. (3), and the internal temperature $T$. When needed, I calculate the effective surface temperature $T_s$ from the internal temperature by the relation derived by Gudmundsson, Pethick, and Epstein (1982). The model includes the variation of the chemical equilibrium state due to spin-down, the relaxation to chemical equilibrium, heating or cooling due to neutrino-emitting reactions, and cooling due to emission of photons from the surface.

---

[2] These two regimes are also discussed more quantitatively, in the context of magnetic field evolution in neutron stars, by Goldreich & Reisenegger (1992).

For cool, degenerate matter close to chemical equilibrium (in the sense that the total energy differs little from its value for perfectly cold matter in chemical equilibrium at the same baryon density), the total energy per baryon can be written as

$$E(n, x, s) = E_0(n) + E_{\text{ch}}(n, x) + E_{\text{th}}(n, s), \tag{6}$$

where $E_0$ is the energy in the cold, chemical equilibrium state of baryon density $n$,

$$E_{\text{ch}} = \frac{1}{2} E_{xx}(x - x_{\text{eq}})^2 = \frac{1}{2} E_{xx}^{-1}(\delta\mu)^2 \tag{7}$$

is the "chemical" energy (due to the departure from chemical equilibrium), and

$$E_{\text{th}} = \frac{1}{2} E_{ss} s^2 = \frac{1}{2} E_{ss}^{-1} T^2 \tag{8}$$

is the thermal energy. The second equality in eqs. (7) and (8) follows from the thermodynamic identities $(\partial E/\partial s)_{n,x} = T$ and $(\partial E/\partial x)_{n,s} = \delta\mu$. The time evolution of the chemical energy is given by

$$\dot{E}_{\text{ch}} = E_{xx}(x - x_{\text{eq}})(\dot{x} - \dot{x}_{\text{eq}}) = E_{xx}^{-1} \delta\mu \dot{\delta\mu}$$
$$= -\delta\mu \left( \alpha n \frac{E_{nx}}{E_{xx}} \frac{\Omega\dot{\Omega}}{G\rho_c} + \frac{\Gamma}{n} \right). \tag{9}$$

In the last expression, the first term in parentheses accounts for the change in the chemical equilibrium state due to spin-down (or spin-up), and the second term for the change in the actual chemical state due to reactions. Similarly, the thermal evolution is determined by

$$\dot{E}_{\text{th}} = E_{ss}^{-1} T\dot{T} = \frac{\Gamma\delta\mu}{n} - \frac{\epsilon}{n} - \dot{E}_\gamma, \tag{10}$$

where the three terms in the last expression are, in the same order, the chemical energy released by the reactions, the energy radiated in neutrinos and antineutrinos, and the energy radiated in photons from the surface. Expressions for $\Gamma\delta\mu$ and $\epsilon$ are given in the Appendix. For the last term, I make the rough estimate

$$\dot{E}_\gamma \approx \frac{m_n}{M_\star} L_\gamma \approx 4.5 \times 10^{-25} T_8^{2.2} \text{ erg s}^{-1}, \tag{11}$$

where $m_\mathrm{n}$ and $M_\star$ are the masses of a neutron and of the star, and the surface photon luminosity, $L_\gamma$, is given in terms of the interior temperature, $T = 10^8 T_8$ K, by the relation of Gudmundsson et al. (1982), claimed to hold to better than 1.5% in the range $0.16 \leq T_8 \leq 10$. For illustration purposes, I assume that it also holds for temperatures beyond the lower limit of this range. Since the lower limit is already lower than the typical current detection threshold, this should not cause serious problems.

For $E_{ss}$, I use the value corresponding to a mixture of non-relativistic neutrons and protons and extremely relativistic electrons, all non-interacting and completely degenerate:

$$E_{ss}^{-1} = \left(\frac{\pi}{3n}\right)^{2/3} \left(\frac{k}{\hbar}\right)^2 m_\mathrm{n} \left[(1-x_\mathrm{eq})^{1/3} + x_\mathrm{eq}^{1/3} + \frac{\hbar}{m_\mathrm{n} c}(3\pi^2 n)^{1/3} x_\mathrm{eq}^{2/3}\right]. \tag{12}$$

In practice, interactions among nucleons are expected to reduce their effective mass relative to $m_\mathrm{n}$, but this effect is uncertain and unlikely to change the result by a large factor, so it is ignored here.

In order to estimate $E_{nx}$ and $E_{xx}$, I use the customary (and very good) approximation (see Prakash, Ainsworth, & Lattimer 1988 for references)

$$E_\mathrm{ch}(n,x) = \tilde{E}(n) + S(n)(1-2x)^2 + \frac{3}{4}\hbar c(3\pi^2 n)^{1/3} x^{4/3}, \tag{13}$$

where $\tilde{E}(n) = -S(n)(1-2x_\mathrm{eq})^2 - (3/4)\hbar c(3\pi^2 n)^{1/3} x_\mathrm{eq}^{4/3}$, $S(n)$ is the so-called "nuclear symmetry energy," and the last term is the contribution of the electrons. From eqs. (6) and (13), I obtain

$$E_{nx} = -\left(q - \frac{1}{3}\right) \hbar c (3\pi^2 x_\mathrm{eq})^{1/3} n^{-2/3}, \tag{14}$$

and

$$E_{xx} = \frac{1}{3}\hbar c(3\pi^2 n)^{1/3} x_\mathrm{eq}^{-2/3} \frac{1+4x_\mathrm{eq}}{1-2x_\mathrm{eq}}, \tag{15}$$

with $q \equiv d\ln S/d\ln n$, and $x_\mathrm{eq}$ given by the condition $E_x(x = x_\mathrm{eq}) = 0$, or equivalently

$$\frac{x_\mathrm{eq}^{1/3}}{1-2x_\mathrm{eq}} = \frac{4S(n)}{\hbar c(3\pi^2 n)^{1/3}}. \tag{16}$$

In what follows, I take $\alpha = 0.73$, $\rho_\mathrm{c} = 3m_\mathrm{n} n_0 = 8.0 \times 10^{14}\,\mathrm{g\,cm^{-3}}$, $n = 2n_0$, and $x_\mathrm{eq} = 1/9$, the latter corresponding to the threshold for direct Urca processes to be allowed (Lattimer et al. 1991). The value of $q$ (see eq. [14]) is particularly uncertain. It can not even be decided with certainty whether $q$ is larger or smaller than $1/3$. This determines the sign of $E_{nx}$ and $dx_\mathrm{eq}/dn$, and therefore of the departure from chemical equilibrium. I assume $q = 2/3$, the value for non-interacting, degenerate nucleons. In addition, I follow Taylor, Manchester, and Lyne (1993) in defining the spin-down power as

$$\mathcal{P} \equiv 10^{36} \mathcal{P}_{36} \,\mathrm{erg\,s^{-1}} \equiv -I\Omega\dot{\Omega}, \tag{17}$$

with an assumed moment of inertia $I \equiv 10^{45}\,\mathrm{g\,cm^2}$. Table 4 of Taylor et al. (1993) shows the observed values of $\mathcal{P}$ for a large number of radio pulsars. I also write $\delta\mu \equiv \pi k \times 10^8 \delta_8 \,\mathrm{K} \approx 27 \delta_8 \,\mathrm{keV}$, and $u \equiv \delta_8/T_8 = \delta\mu/(\pi kT)$. With these conventions, and the rates given in the Appendix, the evolution equations (9) and (10) become

$$\begin{aligned}
\dot{\delta}_8 &= \left(-1.0 \mathcal{P}_{36} - 3.8 \times 10^8 \delta_8 T_8^4 \frac{u^4 + 10u^2 + 17}{17}\right) \,\mathrm{Myr^{-1}}, \\
\dot{T}_8 &= \left(1.7 \times 10^7 T_8^5 \frac{21u^6 + 105u^4 - 357u^2 - 457}{457} - 0.92 T_8^{1.2}\right) \,\mathrm{Myr^{-1}},
\end{aligned} \tag{18}$$

if direct Urca reactions are active, and

$$\begin{aligned}
\dot{\delta}_8 &= \left(-1.0 \mathcal{P}_{36} - 2.5 \delta_8 T_8^6 \frac{3u^6 + 105u^4 + 945u^2 + 1835}{1835}\right) \,\mathrm{Myr^{-1}}, \\
\dot{T}_8 &= \left(0.14 T_8^7 \frac{15u^8 + 420u^6 + 1890u^4 - 7340u^2 - 11513}{11513} - 0.92 T_8^{1.2}\right) \,\mathrm{Myr^{-1}},
\end{aligned} \tag{19}$$

if only modified Urca reactions are allowed. The ratios of the coefficients of different powers of $u$ in each equation are exact in the limit of vanishing $\delta\mu$ and $T$ (see Appendix and Haensel 1992), but all other coefficients in these equations are only rough estimates. Note that a modest departure from chemical equilibrium, $|u| \lesssim 1.5$, *increases* the cooling rate, and only larger departures from chemical equilibrium reduce the cooling rate or may even lead to net heating. For both direct and modified Urca processes, the cooling rate is enhanced by a maximum factor $\approx 1.5$ for $|u| \approx 1.1$.

### 3.3. Results

Neutron stars are born hot ($T \sim 10^{11}$ K), but with rotation rates and magnetic fields not enormously larger than the observed ones. Thus, in the earliest phase of the evolution, spin-down heating and photon cooling are negligible, and only Urca reactions are important. Under these assumptions,

$$\beta \equiv \frac{\dot{\delta\mu}/\delta\mu}{\dot{T}/T} \gtrsim 10 \tag{20}$$

for both direct and modified Urca reactions. This result implies that $|\delta\mu|$ decreases much faster than $T$, and soon only the lowest-order powers of $u$ contribute to the evolution. In this case, $T \propto t^{-1/4}$ for direct Urca reactions, $T \propto t^{-1/6}$ for modified Urca reactions, and $|\delta\mu| \propto t^{-\beta}$ for both. Thus, as has been argued before (e.g., Maxwell 1979), but (to my knowledge) never shown explicitly, a primordial departure from chemical equilibrium in a nonrotating star decays on a timescale substantially shorter than the cooling time, unless Urca reactions are not the dominant cooling mechanism, and the latter does not alter the chemical composition.

In a star that is spinning down, $|\delta\mu|$ increases again after the temperature (and hence the reaction rate) is low enough. If the spin-down power can be approximated as a constant (which is the case at early times for a pulsar with magnetic dipole braking), one has in this stage

$$\delta_8 \approx -\frac{\mathcal{P}_{36}}{1+\beta'}\frac{t}{\text{Myr}}, \qquad (21)$$

with $\beta' = \beta/4 \approx 6$ for direct Urca reactions, and $\beta' = \beta/6 \approx 3$ for modified Urca reactions, according to equations (18) and (19). $|\delta\mu|$ increases in this way until it is of order $kT$, at which point higher order powers of $u$ become important in the expressions for the reaction rates, and the conversion of chemical energy into thermal energy is speeded up. Only after this time does the finite departure from chemical equilibrium substantially affect the thermal evolution.

If $\mathcal{P}$ varies slowly over the timescales required to cool the stellar core and achieve chemical equilibrium, then a quasi-equilibrium state is reached in which the terms on the left-hand sides of equations (9) and (10) are negligible compared to individual terms on the right-hand side, which nearly cancel each other. In this state, the departure from chemical equilibrium and the core and surface temperatures are independent of the previous evolution of the star, and are completely determined by the current value of $\mathcal{P}$, which is directly measurable from pulsar timing, and by the neutron star model (mass, composition, and equation of state).

Fig. 1 shows numerical integrations of equations (18) and (19) together with the corresponding quasi-equilibrium solutions. I assume magnetic dipole braking, $P\dot{P} \approx 10^{-39} B^2$, with $P$ in seconds and the surface magnetic field strength $B$ in gauss, and I assume a fast initial rotation, $P_i = 1\,\text{ms}$, which maximizes the heating due to spin-down at early times, $t \lesssim 10^{39} (P_i[\text{s}]/B[\text{G}])^2$ s. The three panels correspond to three different (constant) magnetic field strengths, (a) $B = 10^{12}$ G, (b) $B = 10^{10}$ G, and (c) $B = 10^8$ G. For the strongest field ($10^{12}$ G, representative of a typical, "classical" radio pulsar), the spin-down occurs so rapidly that the heating is only substantial at early times, when the temperature is still quite high. Thus, the only observable effect is a small bump in the cooling curve, which would be even smaller for a (perhaps more realistic) lower initial rotation period. The heating effect becomes more interesting for the intermediate field ($10^{10}$ G). Here, the evolutionary curves follow the quasi-equilibrium curves very closely after $\delta\mu$ has grown enough for the quasi-equilibrium state to be reached for the first time. In the photon-cooling regime, some deviation is seen, because the reactions are too slow to follow the rapidly dropping equilibrium solutions. For the lowest field ($10^8$ G, representative of a millisecond pulsar), the heating is negligible until around the time when the photon-cooling regime is reached. After this point, however, both the departure from chemical equilibrium and the temperature follow the quasi-equilibrium curves, which remain quite high for about a Hubble time. Of course, the "standard model" for the history of millisecond pulsars (see Bhattacharya & van den Heuvel 1991 for a review) is not as uneventful as the evolution of my $10^8$ G neutron star. The magnetic field may have decayed or the star may have been heated by accretion. Still, unless these events were very recent, the star should have evolved to the quasi-equilibrium state, and its luminosity should be predictable from its spin-down power.

Fig. 2 shows the evolution of the effective surface temperature for $B = 10^{12}$ G (dotted), $10^{11}$ G (short-dashed), $10^{10}$ G (long-dashed), $10^9$ G (short-dashed-dotted), and $10^8$ G (long-dashed-dotted), and in the absence of spin-down heating (solid), for direct (lower curves) and modified (upper curves) Urca reactions. Again, one sees that the effect of spin-down heating can be substantial at late times for stars with weak fields. The effect is quite weak for stars with strong fields, and is further reduced if the initial period is longer than the assumed 1 ms.

### *3.4. Speculations on superfluid effects*

So far, there is no published calculation of the reaction rates in superfluid (i.e., BCS-paired) neutron star matter out of chemical equilibrium. Thus, it is not yet possible to compute the coupled thermal and chemical evolution of a superfluid neutron star, not even in a very simplified way such as that presented in the previous sections for a normal neutron star. In the following paragraph, I make educated guesses about the qualitative changes expected in the evolution if pairing effects were taken into account.

The thermal energy at a given temperature (or, equivalently, the heat capacity) of superfluid particles is substantially reduced if the temperature is much smaller than the transition temperature $T_c$. The chemical energy, however, should be nearly unaffected by the pairing. The reactions involving superfluid particles are substantially slowed if $T \ll T_c$ and $\delta\mu \ll kT_c$. The reduction in the reaction rates makes $|\delta\mu|$ rise faster, and allows a larger fraction of the rotational energy to be stored as chemical energy. For the same reason, photon cooling becomes important at a higher temperature, and the ratio $|u| = |\delta\mu|/\pi kT$ is larger. The larger chemical energy is stored for a longer time, allowing heating to take place until later in the evolution of the star. In the quasi-equilibrium regime, the reduced reaction rate increases the photon luminosity for a given spin-down power, just as in normal matter the equilibrium photon luminosity for the modified Urca processes is substantially higher than that for the much faster direct Urca processes.

These arguments suggest that the effects of spin-down heating are more dramatic in a superfluid neutron star than in the normal star studied more carefully in this paper.

### 4. DISCUSSION

The quantitative results presented in the previous section are based on a very simplified (perhaps oversimplified) model that ignores many features to be expected in a real neutron star. These include a non-trivial spatial structure, where different layers can have different properties, and where diffusion from one layer to another (or its absence) can have an important effect. Related to this is the finite timescale for thermal conduction, which is known to modify the early stages of neutron star cooling (Nomoto & Tsuruta 1987). For many of the physical parameters affecting the evolution, I took representative values which may well be off by a factor $\sim 2$ or more. I also considered only two cooling mechanisms (neutrino emission by direct or modified Urca reactions, and surface photon emission), neglecting several others (see Nomoto & Tsuruta 1987). Finally, superfluidity is not included in the model. Its effects are discussed briefly in §3.4.

Two main pieces of work would be needed in order to produce models that could satisfactorily explore the thermal evolution of different model neutron stars. First, in order to include neutron superfluidity and proton superconductivity, it will be necessary to extend the reaction rate calculations to superfluid neutron star matter at finite temperature and away from chemical equilibrium, possibly building on recent work by Haensel (1992) on normal matter at finite temperature and away from chemical equilibrium, and by Levenfish & Yakovlev (1994b) on superfluid matter at finite temperature and in chemical equilibrium. Second, one should use a code with spatial structure in at least one (radial) dimension, in order to take into account the diffusion of particles and heat and to make a quantitative study of some of the effects pointed out in §3.1. Of course, a rotating star is not spherically symmetric but, as I showed in §2, the variables determining the stellar structure are all constant on one set of surfaces, $\mathcal{S}(M)$. This probably makes a one-dimensional description practical. The code could include all known cooling mechanisms, and the effect of General Relativity, in order to produce predictions which could be compared directly and reliably to observations (to the extent that the latter are reliable as well; see below).

In spite of its obvious shortcomings, the simple model shows the main qualitative effects of the conversion of rotational energy into chemical energy, and finally into heat. Since the rate at which energy is converted at each of the two steps depends on the values of $\delta\mu$ and $T$, it cannot simply be treated as a given source term in the thermal evolution. In this respect, it is very different from and somewhat more subtle than the other heating mechanisms studied so far and mentioned in the Introduction. However, it may be regarded as less speculative than the previously proposed heating mechanisms based on vortex creep or crust cracking, since it involves only physical processes and parameters which affect the thermal evolution even in the absence of heating, and not additional (and very uncertain) parameters such as the pinning strengths of vortex lines to the solid lattice of the crust and the yield strength of the latter. It is often attempted to determine the values of the latter by comparing theoretical models for the thermal evolution to observations (e.g., Cheng et al. 1992; Van Riper et al. 1994). Since all heating mechanisms are most noticeable at late times in the thermal evolution of a neutron star, these studies are subject to confusing different heating mechanisms, and their observational disentanglement does not appear feasible in the near future.

It is clear from the numerical integrations (§3.3.) that the proposed new heating effect is only of small to moderate importance in the early (neutrino-dominated) era of the thermal evolution, but can become quite dramatic in later stages, particularly if the interior of the star is superfluid (see §3.4.). It should be interesting to compare the quasi-equilibrium luminosities predicted for millisecond pulsars to observational data. Unfortunately, it is hard to extract the surface luminosity due to the heat flowing out of the stellar interior ("cooling luminosity") from the observations. First, interstellar absorption modifies the spectra and observed luminosities and is hard to correct for, since the intrinsic spectra are not known (e.g., Becker & Trümper 1993; Danner, Kulkarni, & Thorsett 1994). Secondly, the luminosity is usually (perhaps always) dominated by the additional emission coming from the star's magnetosphere or from external (magnetospheric) heating of the polar caps (see, e.g., Ögelman 1994 for a review). Based on "classical" pulsars, Seward & Wang (1988) and Ögelman (1994) have proposed empirical formulae for the magnetospheric x-ray luminosity in terms of the spin-down power, which predict that the former should indeed dominate over the "cooling luminosity" by a large factor (Fig. 3). However, Danner et al. (1994) have pointed out that the luminosities of some millisecond pulsars (detections or upper limits) are much lower than predicted by the empirical relations of Seward & Wang (1988) and Ögelman (1994), and in fact differ significantly among pulsars with very similar spin-down power. This gives hope that some millisecond pulsars may provide a "window" through which the "cooling luminosity" can be seen.

Fig. 3 shows that, if direct Urca processes are permitted, the luminosity originating from the interior should be completely unobservable. This is not the case, however, if only modified Urca processes are allowed, for which the predicted cooling luminosities come quite close to some of the ROSAT PSPC upper limits of Danner et al. (1994). Of course, in view of the deficiencies of the present model and of the uncertainties in the observations, this result can only be taken to indicate that the proposed effect may well be interesting, and that further theoretical and observational investigations are warranted. The former have already been discussed above, and among the latter the most interesting may be x-ray (or extreme ultraviolet) observations of the millisecond pulsar B1937+21, whose spin-down power is only a factor of two lower than that of B1821-24 (see Fig. 3), and which is probably closer ($\sim 3.6$ kpc compared to $\sim 5.1$ kpc; Taylor et al. 1993, Danner et al. 1994).

## 5. CONCLUSIONS

The spin-down of a neutron star perturbs its internal chemical equilibrium state, leading to reactions and to internal heating. Its effect on the thermal history of stars with strong magnetic fields ($\gtrsim 10^{11}$ G) is not very large, but it becomes substantial for weak-field stars, especially at late times. Old neutron stars reach a quasi-equilibrium state in which heating and cooling balance each other, and (for a given neutron star model) the surface photon luminosity can be predicted from the spin-down power. If only modified Urca reactions can occur, this luminosity lies in the observable range for at least some known millisecond pulsars. More detailed calculations, including the determination of reaction rates for superfluid matter out of chemical equilibrium, are needed in order to make precise predictions for specific stellar models.

I am grateful to Peter Goldreich for many enlightening conversations that led to the initial idea for this paper. I also thank Rudi Danner and Shri Kulkarni for useful information on the observations, and Dong Lai, Jordi Miralda-Escudé, and Fred Rasio for clarifying discussions. This work received financial support from the NSF (grant PHY92-45317) and from the Ambrose Monell Foundation.

# APPENDIX
# REACTION RATE AND NEUTRINO EMISSIVITY

In the limit where $kT$ and $\delta\mu$ are much smaller than the Fermi energies of neutrons, protons, and electrons, and assuming that all particles are "normal" (non-superfluid), Haensel (1992) showed that the emissivities (energy per unit volume per unit time emitted in neutrinos/antineutrinos) can be written as[3]

$$\epsilon_\nu(T, \delta\mu) = \epsilon_{\bar\nu}(T, -\delta\mu) = \frac{1}{2}\epsilon(T, 0)\frac{F(\delta\mu/kT)}{F(0)}, \qquad (22)$$

and the emission rates (number of neutrinos/antineutrinos emitted per unit volume per unit time), as

$$\Gamma_\nu(T, \delta\mu) = \Gamma_{\bar\nu}(T, -\delta\mu) = \frac{\epsilon(T, 0)}{2kT}\frac{G(\delta\mu/kT)}{F(0)}. \qquad (23)$$

Here, $\epsilon(T, \delta\mu) = \epsilon_{\bar\nu} + \epsilon_\nu$ denotes the total emissivity, including neutrinos and antineutrinos,

$$F(x) \equiv \int_0^\infty dy \frac{y^3 P(y-x)}{1 + \exp(y-x)}, \qquad (24)$$

and

$$G(x) \equiv \int_0^\infty dy \frac{y^2 P(y-x)}{1 + \exp(y-x)}. \qquad (25)$$

For the direct Urca processes, $P(y) = P_\mathrm{d}(y) \equiv \pi^2 + y^2$, and for the modified Urca processes, $P(y) = P_\mathrm{m}(y) \equiv 9\pi^4 + 10\pi^2 y^2 + y^4$.

For the present work, only the total emissivity $\epsilon$ and the *net* reaction rate $\Gamma = \Gamma_\nu - \Gamma_{\bar\nu}$ are needed. These are proportional to

$$F^+(x) \equiv F(x) + F(-x) = \int_0^\infty dy \frac{(2y^3 + 6x^2 y)P(y)}{1 + e^y} - \int_0^x dy(y-x)^3 P(y), \qquad (26)$$

---

[3] The present sign convention for $\delta\mu$ is that used by Reisenegger & Goldreich (1992) and Goldreich & Reisenegger (1992), and opposite to that of Sawyer (1989) and Haensel (1992).

and

$$G^-(x) \equiv G(x) - G(-x) = 4x \int_0^\infty dy \frac{yP(y)}{1+e^y} + \int_0^x dy(y-x)^2 P(y), \qquad (27)$$

where I have used the fact that $P(y) = P(-y)$. Using the explicit forms for $P(y)$ given above, and the identity

$$\int_0^\infty dy \frac{y^{2n-1}}{1+e^y} = \left(1 - \frac{1}{2^{2n-1}}\right) \frac{(2\pi)^{2n}|B_{2n}|}{4n} \qquad (28)$$

(Gradshteyn & Ryzhik 1980, eq. 3.411.4), where $B_n$ are the Bernoulli numbers ($B_2 = 1/6$, $B_4 = -1/30$, $B_6 = 1/42$, $B_8 = -1/30$), one can write $F^+(x)$ and $G^-(x)$ explicitly as polynomials. Defining $u = \delta\mu/(\pi kT)$, I obtain

$$\epsilon_{\rm d}(T, \delta\mu) = \epsilon_{\rm d}(T, 0)\left(1 + \frac{1071u^2 + 315u^4 + 21u^6}{457}\right) \qquad (29)$$

and

$$\Gamma_{\rm d}(T, \delta\mu)\delta\mu = \epsilon_{\rm d}(T, 0)\frac{714u^2 + 420u^4 + 42u^6}{457} \qquad (30)$$

for the direct Urca processes, and

$$\epsilon_{\rm m}(T, \delta\mu) = \epsilon_{\rm m}(T, 0)\left(1 + \frac{22020u^2 + 5670u^4 + 420u^6 + 9u^8}{11513}\right) \qquad (31)$$

and

$$\Gamma_{\rm m}(T, \delta\mu)\delta\mu = \epsilon_{\rm m}(T, 0)\frac{14680u^2 + 7560u^4 + 840u^6 + 24u^8}{11513} \qquad (32)$$

for the modified Urca processes.

Throughout this paper, I use the (somewhat uncertain) equilibrium emissivities given by Haensel (1992; references to the original derivations and to critiques are given there),

$$\epsilon_{\rm d}(T, 0) \approx 4.3 \times 10^{21} \left(\frac{x_{\rm eq}n}{n_0}\right)^{1/3} T_8^6 \,\,{\rm erg\,cm^{-3}\,s^{-1}} \qquad (33)$$

and

$$\epsilon_{\rm m}(T, 0) \approx 3.5 \times 10^{13} \left(\frac{x_{\rm eq}n}{n_0}\right)^{1/3} T_8^8 \,\,{\rm erg\,cm^{-3}\,s^{-1}}, \qquad (34)$$

where $n_0 = 0.16\,{\rm fm}^{-3}$ is the normal nuclear baryon density.

# REFERENCES

Alpar, M. A., Anderson, P. W., Pines, D., & Shaham, J. 1984, ApJ, 276, 325

Bahcall, J. N. & Wolf, R. A. 1965, Phys. Rev., 140, B1452

Baym, G. 1981, ApJ, 248, 767

Becker, W., & Trümper, J. 1993, Nature, 365, 528

Bhattacharya, D. & van den Heuvel, E. P. J. 1991, Phys. Rep., 203, 1

Boguta, J. 1981, Phys. Lett. B, 106, 255

Brown, G. E., Kubodera, K., Page, D., & Pizzochero, P. 1988, Phys. Rev. D, 37, 2042

Cheng, K. S., Chau, W. Y., Zhang, J. L., & Chau, H. F. 1992, ApJ, 396, 135

Chiu, H. Y. & Salpeter, E. E. 1964, Phys. Rev. Lett., 12, 413

Cutler, C., Lindblom, L., & Splinter, R. J. 1990, ApJ, 363, 603

Danner, R., Kulkarni, S. R., & Thorsett, S. E. 1994, ApJ, in press

Finzi, A. & Wolf, R. A. 1968, ApJ, 153, 835

Gamov, G. & Schoenberg, M. 1941, Phys. Rev., 59, 539

Gnedin, O. Y. & Yakovlev, D. G. 1994, Astron. Lett., 20, 40

Goldreich, P. & Reisenegger, A. 1992, ApJ, 395, 250

Gourgoulhon, E. & Haensel, P. 1993, A&A, 271, 187

Gradshteyn, I. S., & Ryzhik, I. M. 1980, Table of Integrals, Series, and Products, fourth edition (New York: Academic Press)

Gudmundsson, E. H., Pethick, C. J., & Epstein, R. I. 1982, ApJ, 259, L19

Haensel, P. 1992, A&A, 262, 131

Haensel, P. & Schaeffer, R. 1992, Phys. Rev. D, 45, 4708

Kittel, C. & Kroemer H. 1980, Thermal Physics, second edition (New York: Freeman), chapter 5

Lai, D. 1994, MNRAS, in press

Landau, L. D., & Lifshitz, E. M. 1959, Fluid Mechanics (Oxford: Pergamon), §§15 & 78

Lattimer, J. M., Pethick, C. J., Prakash, M., Haensel, P., 1991, Phys. Rev. Lett. 66, 2701

Levenfish, K. P. & Yakovlev, D. G. 1994a, Astron. Rep., 38, 247

Levenfish, K. P. & Yakovlev, D. G. 1994b, Astron. Lett., 20, 43

Maxwell, O. V. 1979, ApJ, 231, 201

Nomoto, K. & Tsuruta, S. 1987, ApJ, 312, 711

Ögelman, H. 1994, in Lives of the Neutron Stars, NATO ASI series, in press

Pethick, C. J. 1991, in The Structure and Evolution of Neutron Stars, ed. D. Pines, R. Tamagaki, & S. Tsuruta (Redwood City: Addison-Wesley), 115

Pethick, C. J. 1992, Rev. Mod. Phys., 64, 1133

Prakash, M., Ainsworth, T. L., & Lattimer, J. M. 1988, Phys. Rev. Lett., 61, 2518

Reisenegger, A. & Goldreich, P. 1992, ApJ, 395, 240

Sakurai, T. 1969, J. Fluid Mech., 37, 689

Sawyer, R. F. 1989, Phys. Rev. D, 39, 3804; erratum 1989, Phys. Rev. D, 40, 4201

Seward, F. D. & Wang, Z.-R. 1988, ApJ, 332, 199

Shibazaki, N. & Lamb, F. K. 1989, ApJ, 346, 808

Taylor, J. H., Manchester, R. N., & Lyne, A. G. 1993, ApJSS, 88, 529

Thompson, C. & Duncan, R. C. 1993, ApJ, 408, 194

Umeda, H., Shibazaki, N., Nomoto, K., & Tsuruta, S. 1993, ApJ, 408, 186

Urpin, V. A. & Ray, A. 1994, MNRAS, 267, 1000

Van Riper, K. A., Link, B., & Epstein, R. I. 1994, preprint (astro-ph 9404060)

Walin, G. 1969, J. Fluid Mech., 36, 289

Wolf, R. A. 1966, ApJ, 145, 834

**FIGURE LEGENDS:**

**Fig. 1.**— Thermal and chemical evolution of neutron stars. The variables $\delta\mu/\pi k$ (a measure for the departure from chemical equilibrium; short-dashed lines) and $T$ (the internal temperature of the star; long-dashed lines) are plotted logarithmically as functions of time, with direct (lower curves) and modified (upper curves) Urca reactions. The dotted curves are the quasi-equilibrium solutions obtained at each point by requiring that the right-hand sides of the evolutionary equations (18) and (19) vanish. For reference, the temperature in the absence of heating is also shown (solid line). The three panels correspond to magnetic dipole braking with different magnetic field strengths: (a) $B = 10^{12}$ G; (b) $B = 10^{10}$ G; (c) $B = 10^8$ G. The initial spin period is taken to be 1 ms.

**Fig. 2.**— Effective surface temperature as a function of time for stars with direct (lower curves) and modified (upper curves) Urca reactions, with no heating (solid lines) or spin-down heating with magnetic field strengths $B = 10^{12}$ G (dotted), $10^{11}$ G (short-dashed), $10^{10}$ G (long-dashed), $10^9$ G (short-dashed-dotted), and $10^8$ G (long-dashed-dotted). The initial spin period is taken to be 1 ms.

**Fig. 3.**— Surface photon luminosity of a neutron star as a function of its spin-down power. The solid line shows the bolometric luminosity as predicted by the model described in this paper for a star whose core consists of normal neutrons, protons, and electrons, and in which only modified Urca reactions are allowed. The long-dashed line shows the corresponding predicted luminosity in the ROSAT PSPC range (0.1 to 2.4 keV), assuming a blackbody spectrum. The short-dashed curve is a rough prediction for the bolometric luminosity with direct Urca reactions. For comparison, I show the empirical curves of Seward & Wang (1988; dot-dashed) and Ögelman (1994; dotted) for the magnetospheric emission, and the ROSAT PSPC observations of millisecond pulsars from Danner et al. (1994). Triangles indicate upper limits, asterisks detections. The filled triangle is the upper limit and likely detection of PSR B1821-24, corrected for absorption.